# On the rework and development of new Geant4 Cherenkov models

B Đurnić[*], A Potylitsyn[a], A Bogdanov[a,b] and S Gogolev[a]

[a] National Research Tomsk Polytechnic University (TPU), Lenin Avenue 30, Tomsk 634050, Russia

[b] Cancer Research Institute of Tomsk NRMC, Kooperativny Street 5, Tomsk 634009, Russia

Abstract: A recent experiment showed that a Cherenkov radiation spectrum from thin radiators possessing frequency dispersion could transform into quasi-monochromatic spectral lines. Simulations based on the standard Geant4 toolkit could not correctly describe the experimental results because Geant4 uses the traditional Frank-Tamm theory developed for an ideal (infinitely thick) radiator. Because of that, our endeavors in this paper are to present a new way of utilizing Cherenkov radiation using new Gneat4 models, while the paper should be used as a helping text to understand the Cherenkov process in Geant4 and what one can expect out of new models. Now, one can analyze quasi-monochromatic Cherenkov spectral lines using Geant4, which is a significant milestone for investigating the possibilities of this technique.

[*] Corresponding author: zobla96@gmail.com

## 1. Introduction

Cherenkov radiation [1, 2] is a type of radiation characterized by its directivity and is emitted due to the interaction between a relativistic charged particle and the surrounding material. The spatial emission of this radiation can be described by equation [3]

$$\theta_{Ch} = \arccos\left(\frac{1}{\beta \cdot n(\lambda)}\right), \qquad (1)$$

where $\beta$ is the relativistic velocity of the charged particle ($\beta = v/c$; $c$ is the speed of light in a vacuum) and refractive index $n$ is a function of the radiation's wavelength $\lambda$. From equation (1), one can write a condition for the emission of Cherenkov radiation as

$$\beta \cdot n(\lambda) \geq 1. \qquad (2)$$

Now, it is evident that the radiation appears if the relativistic charged particle's velocity is greater than the phase velocity of light in a specific medium. However, equation (1) is applicable only for a particular radiation wavelength $\lambda$, while the radiation itself has a continuous spectrum from soft X-ray region [4] to far infrared [5]. The radiation is used as such in a wide variety of experiments [6-8], i.e., as a broad spectrum of energies. Still, a recent experiment [9] showed that one could measure Cherenkov radiation in the form of quasi-monochromatic spectral lines ("peaks" further). For that purpose, the authors used a thin ($L = 200$ µm) quartz plate possessing frequency dispersion [10]

$$n^2 - 1 = \frac{0.6961663\lambda^2}{\lambda^2 - 0.0684043^2} + \frac{0.4079426\lambda^2}{\lambda^2 - 0.1162414^2} + \frac{0.8974794\lambda^2}{\lambda^2 - 9.896161^2}. \qquad (3)$$

The given refractive index is relatively high, which means one cannot extract Cherenkov radiation from a thin perpendicular plate because of total internal reflections. Such a problem is relatively easily solvable by simply rotating the radiator, as shown in Figure 1. As quartz possesses frequency dispersion, various wavelengths are emitted at different angles with respect to passing charged particles. Thus, Cherenkov radiation can be resolved over wavelengths by letting it move long enough through the surrounding vacuum. Therefore, in the experiment, a detector with a small aperture (to capture a small solid angle of the resolved radiation) of $d = 70$ µm was placed $D = 37.6$ cm from the radiator. As shown in Figure 1, the detector angle was $\theta_{vac} = 59.5°$, while the radiator angle $\psi$ was changed from 22.0 to 24.0°. The used monochromatic electron beam of 855 MeV energy had a beam size of $\sigma = 536$ µm and beam divergence of less than 0.1 mrad. The latter is neglected in the rest of this paper.

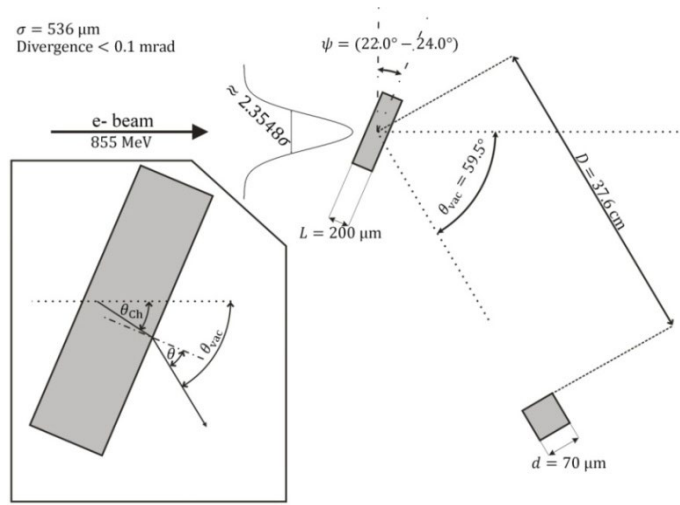

Figure 1 The schematic view of the setup used in the experiment to obtain Cherenkov peaks. The radiation is extracted into a surrounding vacuum and detected by a far-small-aperture detector.

In the following section, we will discuss how to theoretically explain the obtained Cherenkov peaks and provide a few references to the previously obtained results in Geant4. Further, we will point out the limitations of the current Geant4 Cherenkov process and how we have solved them (note that having C++ and Geant4 experience is required to understand this section fully). Throughout the two subsequent sections, we will use and analyze new Geant4 Cherenkov models to understand them better and explain the experimental Cherenkov peaks. Finally, the paper will be finished with a "Summary and Conclusions" section followed by an external link to the Geant4 project with the complete code used in this paper and new Cherenkov processes.

## 2. Theoretical explanation of Cherenkov peaks from thin radiators

The first and the most straightforward approach to theoretically recreate Cherenkov peaks would be to use equation (1). That can be done using the Geant4 toolkit [11–13], as the current version (in this paper, we used version 11.1) is based on equation (1). However, we have already tried it [14, 15] and obtained especially narrow peaks for pencil-like beams. That directly results from equation (1) and the specific experimental geometry. Nevertheless, Geant4 significantly helped discuss and analyze results because more-realistic Gaussian beam (GB) results improved substantially over the pencil-like beam (PB) results. Still, such improvement is directly related to the beam instead of Cherenkov radiation characteristics.

Equation (1) is theoretically derived for an ideal radiator, i.e., an infinitely thick radiator. For such radiators, there is enough space for interference effects of Cherenkov radiation to occur. Because of that, the radiation can be approximated as a delta function, like in equation (1). However, in thin radiators, the traveled distance of emitted Cherenkov radiation is limited, and there is not enough space for interference effects to occur. As a result, for a specific wavelength $\lambda$, Cherenkov cone can be imagined as a cone with a lateral surface of a normal-distribution thickness. Many papers reported that such a distribution depends on $\lambda/L$, or, for instance, in [1]

$$\Delta\theta_{\text{Ch}} \sim \frac{\lambda}{L \sin\theta_{\text{Ch}}}. \tag{4}$$

On the other hand, the equation we will use in this paper can be found in [16]

$$\text{FWHM} \approx \frac{2.78 \cdot \lambda}{\pi L n \sin\theta_{\text{Ch}}}. \tag{5}$$

Let us also define a variable **K** as

$$\mathbf{K} = 2.78/2.355\pi \approx 0.38, \tag{6}$$

where the 2.355 is taken to include a Gaussian distribution sigma value. Please note that in the past, we cited [17] as a source of equation (5), which is incorrect. Also, in the equation we reported before, we did not include refractive index $n$, like in equation (5)! Still, due to its simplicity, equation (5) is perfect for use in Monte Carlo simulations (it is not processor-heavy) as we did in this paper.

Let us remember another theoretical method, the "Polarization Currents Method" (PCM). The method was thoroughly discussed elsewhere (e.g., see [18–20]), and it was used to explain the experimental results [9]. Moreover, after understanding the importance of including real-beam characteristics in theoretical methods, we tested our understanding numerically by using real-beam characteristics with the PCM [21] (further RBPCM). The results showed a better agreement with the experimental data and will be used besides the PCM for comparing purposes in this paper.

## 3. Introducing new Geant4 Cherenkov models

Before approaching the problem of how the new Cherenkov models were prepared, we will introduce some additional limitations of the current Geant4 Cherenkov process (class `G4Cerenkov`) throughout the following subsection. Also, note that we encourage the reader to follow the complete code (see the "External link" section) while reading explanations in this section.

*3.1. Refractive index and physics tables in the current Geant4 Cherenkov process*

From the discussion provided in the previous section, it is evident that the current Geant4 Cherenkov process should not be used to simulate Cherenkov radiation in thin radiators. However, we noticed another limitation in the `G4Cerenkov` class while analyzing the source code (also reported in [22]). As the `G4Cerenkov` class is written according to Frank-Tamm theory (see [3]), the number of emitted photons per traveled unit length can be expressed as

$$\frac{dN}{dl} = \frac{2\pi\alpha z^2}{hc}\int_{\varepsilon_{min}}^{\varepsilon_{max}}\left(1 - \frac{1}{\beta^2 n^2(\varepsilon)}\right)d\varepsilon, \quad (7)$$

where $\alpha$ is the fine-structure constant (1/137), $z$ is the charge of a passing charged particle, $h$ is the Planck constant, and $\varepsilon$ is the energy of emitted Cherenkov photons. Here, the authors of the `G4Cerenkov` class had properly noticed physics tables could be built around equation (7) to speed up calculations and considered:

$$\int_{\varepsilon_{min}}^{\varepsilon_{max}}\left(1 - \frac{1}{\beta^2 n^2(\varepsilon)}\right)d\varepsilon = \int_{\varepsilon_{min}}^{\varepsilon_{max}}d\varepsilon - \int_{\varepsilon_{min}}^{\varepsilon_{max}}\frac{1}{\beta^2 n^2(\varepsilon)}d\varepsilon. \quad (8)$$

In `G4Cerenkov`, the physics tables are calculated in the pre-run time for the right integral, i.e., based on $d\varepsilon/n^2$. Further, those physics tables are used in the `GetAverageNumberOfPhotons` method, where maximal and minimal refractive index values are obtained through the `GetMinValue` and `GetMaxValue` methods of the `G4PhysicsVector` class, respectively. However, suppose one checks the definitions of the previous methods. Then, one will notice they only return the first and last values of registered refractive indices (note that the methods would be called `GetFrontValue` and `GetBackValue`, respectively, according to the STL naming convention). That means the `G4Cerenkov` process can be used adequately for most materials as they possess refractive index dependency, as shown in Figure 2 (a). The arrow in the figure shows how the `G4Cerenkov` class searches for the proper refractive index when Cherenkov photons cannot be generated on the whole loaded spectrum.

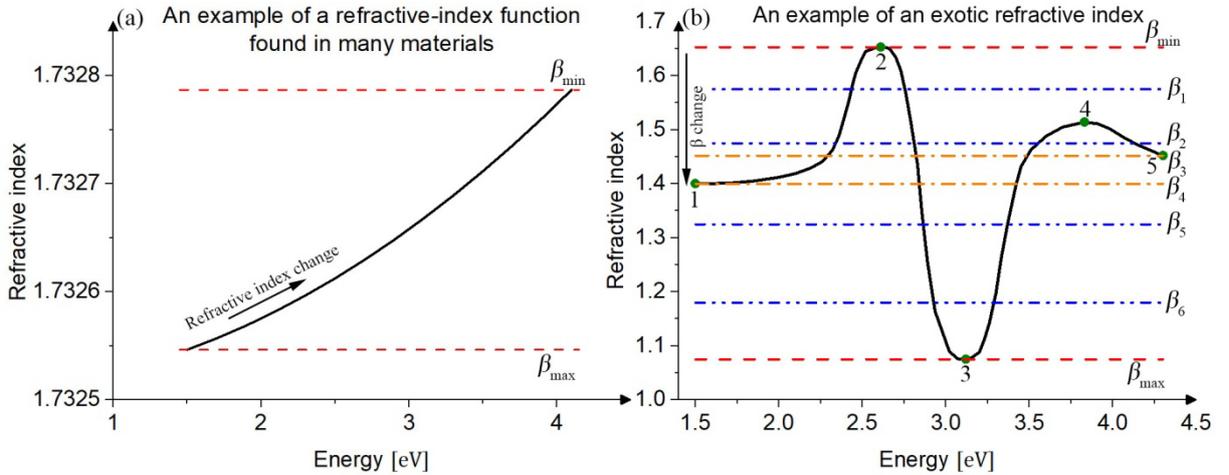

Figure 2 Two examples of arbitrary refractive index dependencies. The standard Geant4 Cherenkov process can be used appropriately for dependencies, such as in plot (a), while it cannot be used for plot (b).

However, the next question is what happens for exotic refractive indices, as shown in Figure 2 (b). According to the previously explained, Geant4 cannot take into account such dependencies, and it will consider that the minimal refractive index is at point 1 and the maximal at point 5. That means if equation ([2](#)) is satisfied for points 2 and 4, Cherenkov photons will not be emitted unless it is also satisfied for point 5. Moreover, point 3 can reduce the number of emitted Cherenkov photons and eventually lead to a negative number of emitted photons (undefined behavior) because Geant4 regards point 1 as a minimal refractive index.

To solve the problem, let us draw horizontal lines based on equation ([2](#)), as shown in Figure [2](#) (b). Now, instead of increasing the $\varepsilon_{min}$ integral limit (remember the arrow in Figure [2](#) (a); also, that approach cannot consider more exotic refractive index dependencies) in equation ([8](#)), we can calculate both sides of the integral based on the relativistic charged particle's beta ($\beta$) value (see the arrow in Figure [2](#) (b)). Of course, those calculations are done depending on the particle's $\beta$ value relative to the horizontal lines (the part of the refractive index (thick black) line above an equation ([2](#)) condition line that can be drawn for the particular $\beta$ value). Nevertheless, besides solving the problem of the number of emitted Cherenkov photons, we also need to generate their energies with a proper distribution. Still, such a problem is much more demanding from a programming point of view, and one should follow the [External link](#) section to understand how it was solved (we will also provide some distribution results in the following section, which will help the reader to understand the code logic).

*3.2. The external code and new Geant4 Cherenkov models*

First and foremost, let us consider the structure of the new (external) code. It is divided into two targets, i.e., the library and executable targets. Here, we will consider only the former target that is statically linked to the latter (a standard Geant4 application) and contains the full information about the new Cherenkov models. As we believe the code might be added in an official Geant4 release, the library has already been written according to the Geant4 naming convention.

The first problem that must be solved is creating a daughter of the `G4VPhysicsConstructor` class, through which one may register a physics in a `G4VModularPhysicsList` object. In current Geant4, such a class is called `G4OpticalPhysics`. However, that class has no virtual methods (including destructor), and there is no point in inheriting it. Because of that, we introduced a new physics class, namely `G4OpticalPhysics_option1`, which has virtual methods and loads all of the optical processes separately. That means one can now override a single virtual method to load a desired user optical process without rewriting the whole physics. That was done in the class `G4OpticalPhysics_option2`, which loads another new Cherenkov process. Note that except for the Cherenkov process, all loaded processes of `G4OpticalPhysics_option1` are the same as `G4OpticalPhysics` processes (they are just instantiated differently).

If one takes a look at the `G4Cerenkov` class that works correctly for most users' needs, one can also understand that the process is relatively simple, and no matter how the new processes are written, we cannot expect to increase the code performance, i.e., we can expect some extra number of processor cycles. Therefore, one is encouraged to use a standard model based on equation ([1](#)) whenever possible. Because of that, a new `G4StandardChR_Model` class has been added. This class is similar to the `G4Cerenkov` class, while the new physics tables are used (see subsection [3.1.](#) and section [4.](#)). However, we also need other models that can replace the standard model when it cannot be adequately utilized. Therefore, a `G4CherenkovProcess` class (loaded through `G4OpticalPhysics_option2`) serves as a wrapper for various Cherenkov radiation models (their abstract class is `G4BaseChR_Model`) and helps in selecting what Cherenkov model should be executed. The selection is done based on the logical volume in which a particle is located. A `std::unordered_map<const G4LogicalVolume*, neededData>` was used for that purpose. Furthermore, the approach was chosen because we also need to efficiently (even if parametrized volumes are used) access the material thickness information required for a finite-thickness Cherenkov model (`G4ThinTargetChR_Model`). This Cherenkov model is based on equation ([5](#)) and will be tested further in this paper.

As the standard model should be used most of the time, except for the `G4CherenkovProcess` class, we provide another Cherenkov process – `G4StandardCherenkovProcess`. This process is loaded through the class `G4OpticalPhysics_option1` and is based on the `G4StandardChR_Model` class (with new physics tables). That is done because it can slightly improve the code performance for very complex detectors (with a significant number of logical volumes) when a user does not desire to use additional, other than standard Cherenkov models. Nevertheless, if only materials with refractive indices, as in Figure 2 (a), are used, and only the standard model is needed, the `G4Cerenkov` class will give as good results as the `G4StandardCherenkovProcess` class!

We finalize the new Geant4 Cherenkov code analysis here because this section aims to explain the basic logic behind the code. Moreover, presenting the entire code would be cumbersome and is not considered here. Again, note that one can find the complete C++ code (currently version 0.5) by following the provided link in the "External link" section at the end of this paper.

### 4. The code performance for exotic refractive indices

In Figure 2 (b), we provided an example of an exotic refractive index and said that the standard Geant4 cannot analyze it properly. Now, let us understand how the new physics tables approach the problem of generating Cherenkov photons for exotic refractive indices. As we do not want processor-heavy code, the generation process is approached from the CDF (cumulative density function) point of view. The corresponding CDF is determined based on the relativistic $\beta$ value of a passing charged particle, and such distributions (from the code) are provided in Figure 3 (a). All those distributions are generated based on the refractive index function shown in Figure 2 (b). Finally, in Figure 3 (b), we provide complete photon spectral distributions produced using the corresponding CDF values.

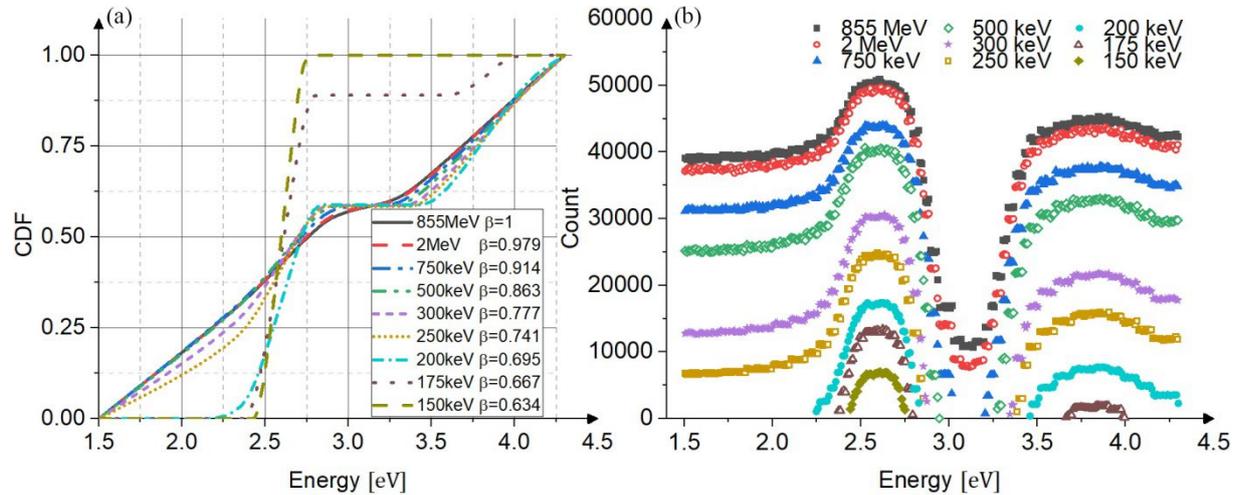

Figure 3 (a) CDF distributions generated for various energies (relativistic $\beta$ values) of electrons and refractive index dependency from Figure 2 (b). (b) Energy distributions of Cherenkov photons produced by electrons of various energies.

Note that all the provided energies of photons are considered at their production point. Also, in places where a decrease in refractive index is present (such as in point 3 in Figure 2 (b)), we can expect a significant increase in the imaginary part of the refractive index, but that is another process we did not consider here (we consider only generation of Cherenkov photons). Still, in Figure 3 (b), one can notice two distinguished peaks in the formed spectra related to points 2 and 4 in Figure 2 (b). In contrast, we can also notice sudden breaks in the spectra between the two peaks because of point 3 in Figure 2 (b), for which equation (2) is not satisfied.

We, the authors of this paper, are unaware of an experiment using Cherenkov radiation and confirming distributions for such exotic refractive indices in the visible region. That means the results are based on equation (7) and should be experimentally confirmed. On the other hand, such peaks in the emitted Cherenkov radiation can be found in the X-ray region, e.g., [4]. Still, as we have not focused on solving the problem of Cherenkov radiation in the X-ray region, the current code cannot produce gamma photons. On the other hand, it is possible that the problem might be solved just by changing the produced particle from "opticalphoton" to "gamma" for corresponding energy, but that is something we should consider in the future.

## 5. Using the modified Geant4 to obtain experimental Cherenkov peaks

Let us start this section by providing the first Cherenkov peaks obtained using the modified Geant4, as in Figure 4. In the provided plots, one can observe that even the PB gave broad peaks and that the extremely narrow peaks reported in [14, 15] are not present anymore. Still, the obtained PB and GB Cherenkov peaks were slightly wider than the PCM ones. Because of that, we have decided to slightly adjust the estimation of the variable **K** from equation (6), i.e., we have reduced it from the original 0.38 to 0.3485, the value that will be considered in the remaining part of this section. Note that we did it because the described experiment entirely depends on the angular distribution of Cherenkov radiation emitted in thin radiators, meaning it can be used for its investigation. Still, we are currently limited to only five experimental Cherenkov peaks, which might be a source of inaccuracy. Therefore, the variable **K** can be modified using a UI command in the provided code. As before, the spectral difference in the location of the experimental and theoretical Cherenkov peaks can be attributed to equation (3), i.e., the refractive index of the quartz was not measured in the experiment but taken as a standard.

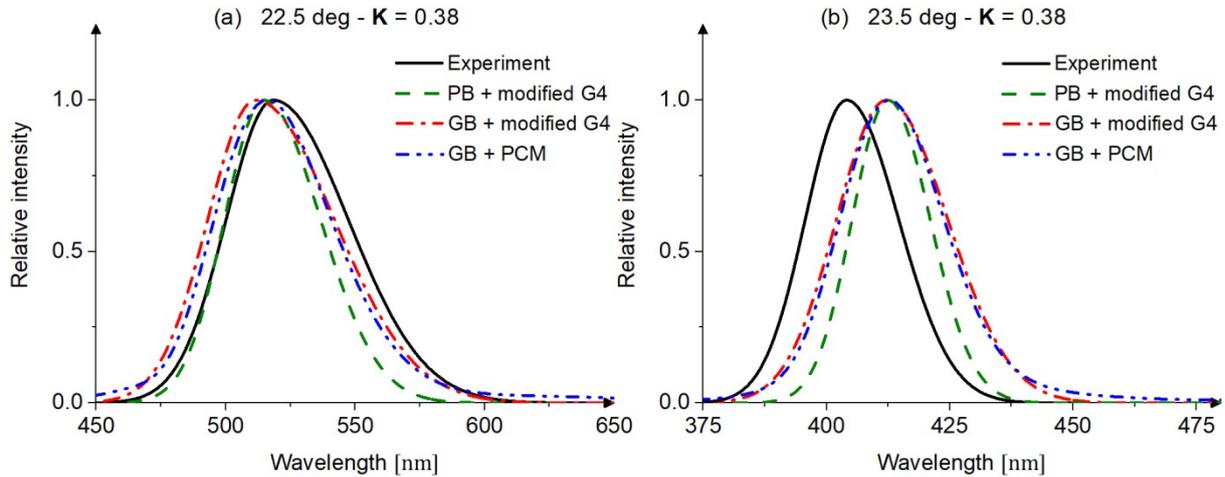

Figure 4 The comparison of the first Cherenkov peaks, obtained using the modified Geant4, with the experimental and theoretical (RBPCM) ones.

Now, let us consider all the spectra of Cherenkov peaks obtained using the modified Geant4, as shown in Figure 5. After adjusting the variable **K**, the similarity between the PCM and modified Geant4 is unsurprising (it was our goal). That is well observed in Figures 5 (a) and (b). Some slight differences between the two can be attributed to relatively low statistics of the Cherenkov peaks obtained using modified Geant4. It is also interesting to point out that the relative intensities of the modified Geant4 instantly changed to the PCM levels because we included Cherenkov radiation angular distribution [14] (the meaning of equation (4)). As in the past, we still leave the question of differences between the theoretical and experimental relative intensities open until we measure some additional experimental Cherenkov peaks.

In Figure 6, similar to [21], we also compare the FWHM values of various peaks. Here, we can observe that the modified Geant4 model gives seemingly statistically identical FWHM values with the PCM. Because of that, we additionally provide all the FWHM values in Table 1. As we observed in Figure 6, the differences are minimal, and we can consider that the variable **K** of 0.3485 describes the experimental results well enough, at least until we measure some additional Cherenkov peaks. Also, we can notice another reason for adjusting the variable **K**, i.e., the theoretically obtained Cherenkov peaks for high radiator inclination angles in the GB case showed somewhat higher FWHM values than the experimental values.

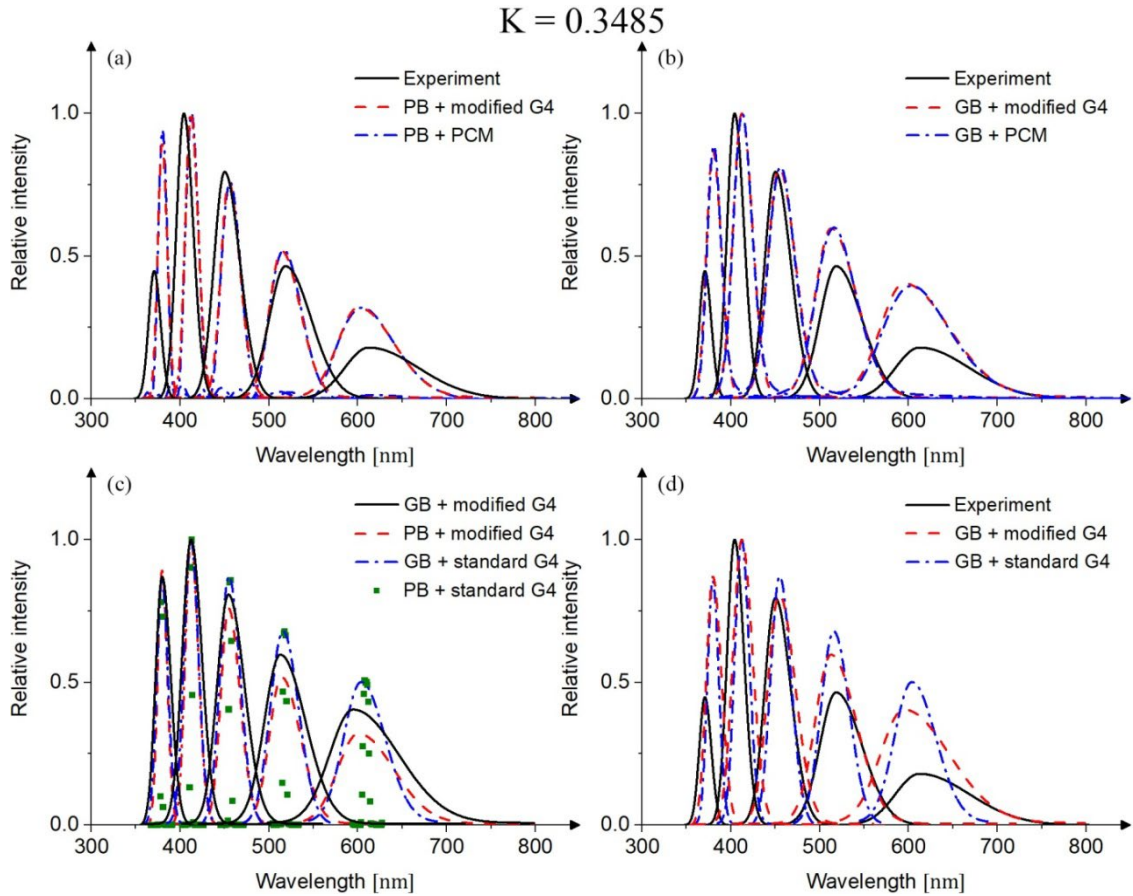

Figure 5 The comparison of obtained spectra for (a) PB (pencil-like beams); (b) GB (Gaussian beams); (c) standard and modified Geant4 models; and (d) standard and modified Geant4 models with the experimental spectra.

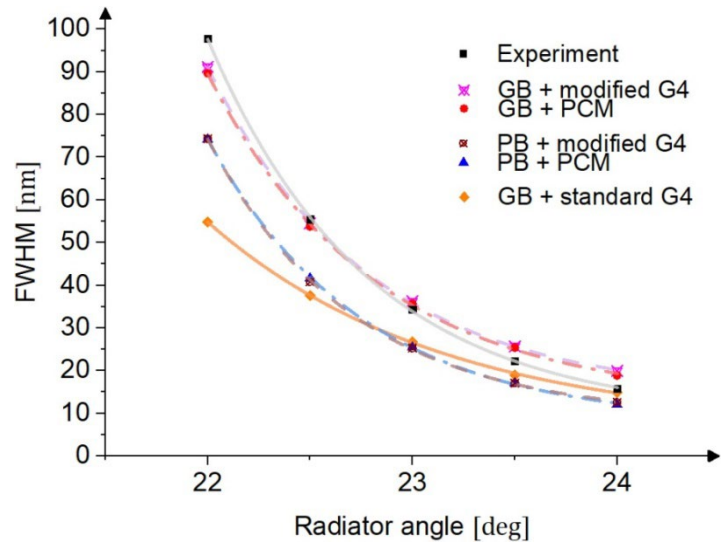

Figure 6 The dependency of the FWHM on the radiator angle for various approaches.

Table 1 The FWHM results obtained from Cherenkov peaks using various approaches (GB = Gaussian beam; PB = pencil-like beam).

| Radiator angle [°] | FWHM [nm] PB + PCM | FWHM [nm] GB + PCM | FWHM [nm] GB + standard G4 | FWHM [nm] Experiment | FWHM [nm] PB + modified G4 | FWHM [nm] GB + modified G4 |
|---|---|---|---|---|---|---|
| 22.0 | 74.23 | 89.65 | 54.82 | 97.74 | 74.35 | 91.15 |
| 22.5 | 41.67 | 53.77 | 37.58 | 55.48 | 40.84 | 54.50 |
| 23.0 | 25.48 | 35.67 | 26.80 | 34.38 | 25.30 | 36.28 |
| 23.5 | 17.10 | 25.39 | 19.00 | 22.29 | 17.11 | 25.68 |
| 24.0 | 12.08 | 18.87 | 14.93 | 15.83 | 12.56 | 19.94 |

## 6. Summary and conclusions

The paper provides many details about how the current Geant4 Cherenkov process works, in what cases it can be used, and when it cannot. We also explain how to improve and generalize the Geant4 Cherenkov model and provide new models to solve various problems in different study cases. Currently, we have provided two Cherenkov radiation models – one based on the standard Frank-Tamm theory (`G4StandardChR_Model`) and the other that can consider finite-thickness radiators (`G4ThinTargetChR_Model`). Note that the latter model can only be used when a charged particle crosses a plate with the thickness *L*, almost perpendicularly to the surface (see Figure 1). Also, it considers that only a single dimension is finite, while the transverse dimensions are infinite. Both provided models use newly written physics tables that allow any refractive index dependencies of materials. Note that we assume a similar can be used for generating Cherenkov radiation in the X-ray region [4] (it is needed to change from `opticalphoton`s to `gamma`s), but we leave the problem open as no particular attention was yet devoted to it.

Besides providing new insights into Geant4 Cherenkov models, we also test them. We have used an arbitrary exotic refractive index dependency to test new physics tables. Still, while Cherenkov radiation is validated for the X-ray region, we are unaware of an experiment with such a refractive-index dependency in the optical region, i.e., the predicted distributions should still be experimentally confirmed. Also, the modified Geant4 could explain experimentally obtained Cherenkov peaks, and after some adjustment, it was comparable to the PCM results. That will be particularly important for us if we should use the technique for beam diagnostics, as proposed in [21, 23].

Finally, we can say that this paper can serve as a helping text to understand the Geant4 Cherenkov process and how it can be used. Nevertheless, we should point out that even the new models have limitations. That means one still cannot use Geant4 to generate Cherenkov photons in volumes of more complex shapes and small sizes or generate the so-called Cherenkov diffraction radiation [5]. Still, we can consider the presented results as a first step towards including more realistic Cherenkov radiation models in Geant4.

**External link**

The authors of this paper believe in free science and that science should be available to everyone. Moreover, we used a free toolkit (Geant4) to obtain the results presented in this paper. Therefore, we are thrilled to announce that the newly developed code has been uploaded as open-source and can be found on a GitHub account, zobla96/ChR_project.

**Acknowledgments**

This research was supported (partly) by the Russian Ministry of Science and Higher Education, project No. FSWW – 2023–0003.